\documentclass[]{aastex63a}

\usepackage{natbib}

\begin{document}

\title{New evidence on the lost giant Chinguetti meteorite}

\author{Robert Warren}
\affiliation{40 Mill Rd, Salisbury, SP2 7RZ, UK}
\author[0000-0003-1599-5402]{Stephen Warren}
\affiliation{Astrophysics Group, Imperial College London, Blackett
  Laboratory, Prince Consort Road, London SW7 2AZ, UK {\em s.j.warren@imperal.ac.uk}}
\author{Ekaterini Protopapa}
\affiliation{Department of Physics, University of Oxford, Clarendon Laboratory, Parks Road, Oxford, OX1 3PU, UK}
\date{\today}

\begin{abstract}
The giant Chinguetti meteorite that Gaston Ripert reported seeing in 1916 has never been found. A radionuclide analysis by Welten et al (2001) of the 4.5kg mesosiderite that Ripert recovered, supposedly sitting on the larger object, has convinced many that Ripert was mistaken, and interest in the giant meteorite has subsequently faded. Aspects of Ripert's account of the giant meteorite are nevertheless compelling, particularly his description of ductile metal needles in one area of the surface explored. Several visual searches for the giant meteorite, beginning in 1924, might have failed because the object was already by then covered in sand. Using DEM data we have measured dune heights and established their drift speed. This has allowed us to create a map of locations where the meteorite could lie. The 2004 PRISM-I aeromagnetics surveys, acquired by Fugro for the Mauritanian Government for Mining Sector capacity building purposes, have the necessary area coverage, spatial resolution, and sensitivity to establish if the meteorite exists. In Jan 2023 we requested the PRISM data from the Ministry of Petroleum Energy and Mines, explaining, under Confidentiality, the scientific purpose of the request. To date the data have not been made available to us. 
\end{abstract}


\section{Brief history and rationale}

This paper concerns the Chinguetti meteorite, a huge 40m-high, 100m-long iron inselberg located in the remote high dunes of the Mauritanian desert, that a French officer, Captain Gaston Ripert, claimed to have seen in 1916. His story reads like a fantasy, complete with an overheard conversation between camel drivers, a furtive mission at night, guided by the local chief, who forbade Ripert to take a compass, and was later poisoned, and a meteorite larger than any known by orders of magnitude, but without an impact crater; and yet Ripert produced a 4.5 kg meteorite he said he found on top \---\ a very interesting object in its own right \---\ and his matter of fact account, his proven scientific expertise, his sincerity, and above all some scientific details in his description of the huge meteorite cannot be reconciled unless the meteorite exists. Multiple searches stretching over 75 years have failed to locate it. The Chinguetti meteorite remains a fascinating mystery that has never been satisfactorily disproved. We are by no means the first to note that what is needed is a magnetometer survey over the entire region where the meteorite could possibly lie.

A comprehensive history of the topic is provided in the book Le Fer de Dieu \citep{FerdeDieu}, in French, and most of the details in the brief summary presented in this section are from this source. An excellent account in English, based primarily on the book, and including additional details is given by \citet{Marvin2007}. It must be appreciated that the region is extremely remote and was largely unmapped when searches for the meteorite commenced in 1924. Also, communication then was by surface mail (Ripert had moved to Cameroon). Because the mission was undertaken at night, without compass, Ripert could provide few details of the location, save that the trip on camelback took some 10h. A miscommunication meant that searches initially concentrated in the region SW of Chinguetti, but later were extended to all areas to the south. Several searches were undertaken by Th\'{e}odore Monod starting in 1934 that continued intermittently until the 1980s, and there was one final search by Phil Bland and Sara Russell in the late 1990s, the subject of a Channel 4 documentary. With the exception of the last, these were all visual searches either on camelback or using aerial photography. (In truth in the 1950s the French army employed a crude magnetic device, a declinometer, in a search. Details are sparse, few measurements were made, and the sensitivity of the instrument is not stated.) Ripert described the meteorite as  being nearly covered by sand in 1916, and so it could have become buried and thereby been missed by all the visual searches. Bland and Russell carried a magnetometer but took only a few measurements (with no positive results) at one particular location, directed there by a pilot Jacques Gallou\'{e}dec who had sighted a feature he hoped could be the meteorite from his plane. Monod also went to the location, but found nothing. We suggest that a shale diapir, an ephemeral phenomenon which we have observed a number of times in the area, is a more likely explanation of what Gallou\'{e}dec saw. These blocky dark masses stand out strikingly, silhouetted against the light-colored dunes, and are hard to fathom until seen up close.

To summarize: it is possible that the meteorite became covered by sand within a few years, and because the initial searches were in the wrong direction, it is conceivable that the meteorite was missed, and remains hidden in the high dunes, still waiting to be discovered.

We now provide a very brief summary of the arguments for and against the existence of the meteorite. The lack of an impact crater is hard to understand, but \citet{Thomas1951} showed it might be explained by the flight of the meteorite being nearly tangential to the surface of the Earth. A study by \citet{Welten2001} of the small meteorite provides the strongest evidence against the existence of the large meteorite. An analysis of radionuclide data indicates (assuming spherical symmetry) that the small meteorite could not have belonged to a parent mass larger than 1.6m in diameter, and the authors suggest Ripert was mistaken, or lied. 

Jean Bosler, Professor of Astronomy at Marseille, invited Ripert to visit him and grilled him on his story. He was greatly impressed by Ripert's sincerity, and noted that if he had invented his account, he would hardly be eager to be interviewed. The most convincing evidence for the large meteorite is Ripert's description of metallic needles in one area, that he tried to break off by hitting with the small meteorite, but they proved too ductile and he could not obtain a sample. As detailed in the paper by \citet{Marvin2007}, in 2003 William Cassidy reported the existence in iron meteorites of spikes of Ni rich material that show similar ductile behavior \---\ this information (both the spikes and their ductility) was unknown to anyone in the scientific community in 1916, let alone Ripert. Indeed, if Ripert's evidence is believed, then he holds the scientific priority on this observation. So there are compelling arguments both for and against the existence of the large meteorite. This provides the rationale for a magnetics survey of the region.

\section{Considerations for a magnetometer survey}

 In 2020 we began a search for the meteorite by collecting and analyzing all available relevant remote-sensing data covering the region. No doubt people have searched for the meteorite in Google Earth (as we have) but we believe we are the first team to apply remote sensing data in an extensive way to this problem. The most useful resources are  SRTM and ALOS digital elevation model (DEM) data. We also analyzed PALSAR radar and Landsat thermal data to supplement this information. For the radar data we co-added images from several epochs to enhance the signal-to-noise ratio, hoping to detect half submerged rocks in areas of interest. For the Landsat data, which are daily images, we searched for points where the temperature variation over the year differed compared to the temperature variation of the surroundings. However nothing of great interest came out of these analyses and the data are not mentioned further.

Since Ripert described the meteorite as 40m high but visual searches have failed to locate it, it could only be hidden in regions of high dunes. Two crucial pieces of information therefore are the radial distance from Chinguetti a camel could reasonably travel in 10h in the terrain encountered south of Chinguetti, and a map of the high dunes within that radius. There are two areas of high dunes in the vicinity of Chinguetti, and they are visible in Fig.1 as the stripey regions (the exact manner in which the stripes were drawn is described below). Immediately to the south of Chinguetti is the band of dunes known at Les Boucles. Nearly all these dunes lie within a radius of 20\,km of Chinguetti. There is a further region of high dunes in the Batraz (Battrass) area between 40\,km and 60\,km to the SE. There is nowhere else for the meteorite to hide.

\subsection{The distance from Chinguetti}

The direction of the meteorite from the starting point of Chinguetti is unknown. At first Ripert declared that they went to the SW, but later said that he meant it was to the SE. He also said that they made detours on the way. In the dark, without a compass, and along a disorienting route, it is safest to assume that he did not know where they finished up relative to Chinguetti. He states the trip was made `en aveugle'. It is unclear whether this translates as `blind' i.e. without compass, or literally blindfolded as some interpret.

As for the distance that they traveled, we can put an accurate upper bound on this. We have made two excursions into the same area south of Chinguetti with highly experienced chameliers from the district. The first trip lasted 11 days, and the second 6 days. The daily average speeds varied from 2.6 km/hr to 3.9 km/hr, when the camels were burdened. These numbers exclude resting time. When unburdened, the average speed rose to 5.0 km/hr. This figure is based on a long round trip one day, across generally good terrain, including the fast inter-dune ground where the camels trotted, as well as briefly traversing a saddle in one large dune in both directions. This again excludes rest time. We can guess that Ripert and his guide would have been largely unburdened, with just enough food and water for a day or two at most. 

However, there are a number of reasons why an average of 5.0 km/hr for 10 hours unbroken, and all in a straight line, cannot be achieved in practice. We interviewed the chameliers (some of them third-generation), and pressed them hard on the maximum distance that could be achieved in a day unburdened. Their primary concern in answering our questions was always the welfare of their camels, which are not only their primary asset, but also their lifeline to returning safely. On our trips they never traveled more than 4 hours in a stretch, and invariably took a 3-hour break in the middle of the journey so that the animals could feed to gain sustenance for the second leg. The thought that Ripert traveled for 10 hours without a break is therefore unrealistic.

Additionally the possibility to travel at an average speed of 5 km/hr requires mostly smooth terrain, or else the risk of a fall is too great. These smooth areas do exist in the inter-dune regions (seen darker on satellite images), and in a prominent  ancient river bed `Taiaret Idaouali', now filled, to the south of Les Boucles, but they are restricted in extent. Furthermore, once away from the smooth areas, the chameliers do not take their camels in a straight line. Within areas of small dunes, we saw them send out a guide on foot ahead to lead them on a zig-zag route avoiding any steps and ledges. And on a larger scale, they always preferred to avoid going over the top of the large dunes, despite our strenuous attempts to get them to change their minds. The zig-zag nature of travel at a small scale, and the avoidance of going over large dunes, means that a day's journey is never in a straight line, and since Ripert traveled at night, these cautions of basic safety must have applied to his journey. On our journeys, the straight-line distance as a proportion of distance traveled on the ground varied from 78\% to 95\%. 

Bearing in mind also that Ripert states that they took a number of detours on the way to their destination, we can say that 50km in a straight line from Chinguetti is an absolute upper limit for how far they could have traveled overnight. To end up in high dunes, the guide may have taken them rather quickly towards Batraz, along the route of the former river bed, and perhaps they succeeded in reaching the fringes of the dune field there. While it cannot be ruled out, we consider this rather unlikely, based on the discussion above (we followed this route in reverse, and it took us three days with burdened camels). Alternatively the guide  took Ripert on a slower journey through Les Boucles, possibly traversing some of the high dunes, either meandering deliberately to ensure Ripert could not find the place again, or because he wasn't confident of the route. We consider this more likely. Furthermore, it does not seem possible that the direct distance could be less than 10\,km from Chinguetti. Ripert was in charge of the camel corps in Chinguetti and one would think he would have recognized the dunes close to the town.  

Circles of radii 10\,km and 50\,km are marked on Fig. 1 and  we contend that the high dunes contained within this annulus cover all reasonable possibilities for the location of the meteorite.

\subsection{Footprint of the high dunes, and dune speed of travel}

 The DEM data are useful in two ways: firstly to identify which dunes are tall enough to hide the meteorite, and secondly to determine how fast the dunes are traveling. The latter allows us to place restrictions on how far into the dune the meteorite would now be sitting. 

 Ripert stated that the large meteorite mass that he saw was covered on its northeastern side by a large dune. He would have been well-oriented on direction because he made his observations at dawn. Sixteen years later he said he thought it possible that the meteorite could have already been completely covered by sand. The presence of a strong prevailing wind towards the southwest is immediately obvious to anyone who visits the area, and this is what causes the dunes’ migration in that direction. From this information we know that the top of the meteorite mass was at a level 40 m (by Ripert’s estimate) up the southwestern (leeward) slope of a major dune whose crest was greater than the height of the meteorite. Dunes retain their fundamental profile as they migrate, so only dunes today that are greater than the meteorite height are candidates for the dune (now moved) that Ripert was looking at.

 The inter-dune areas are flat at a first level of approximation and thereby provide a datum against which we can measure the height of the dunes. To find all the points on all the dunes that are greater than 40m above the datum, we assess each pixel in the DEM map and take a line a chosen search distance directly to the west. If any pixel along that line is 40m lower than our assessment pixel, then we toggle our assessment pixel on. Otherwise it is toggled off. We need to make the search distance quite long because we will assess pixels up the Western side of a dune, up to the crest, and then down the Eastern side until we are below 40m above datum again. To be confident that we were not losing any points, we settled after trial and error on a search distance of 800m.

We were also concerned that we might miss some feasible hiding places for the meteorite if we took Ripert’s estimate of 40m in height completely literally. So, as a conservative measure, we lowered this restriction to 30m, which fattens up and lengthens the patches.

By this means we created the footprint shown shaded beige in Fig. 1. The meteorite cannot lie outside these shaded bands. Shaded bands in the rocky regions to the west indicate rock cliffs and may be ignored.

In fact, we can tie down the possible location of the meteorite further within these shaded areas. We show below that the dunes can have moved no more than $\sim100$\,m since 1916. This means that if you take the shaded areas and shift them 100m to the NE you will be looking at the dune areas higher than 30m as they stood in 1916. Since the meteorite was only just exposed at that time, the southwestern edge of these shifted shaded areas would mark the possible locations of the cliff face that Ripert was looking at. Therefore, on Figure 1, 100m northeast of the western edge of the shaded stripes marks approximately where the front of the meteorite would be positioned today, or a bit further back if the meteorite had been holding the dune back in that region. 

This would be very important for a ground-based magnetometer survey. We know that 30-40m up the leeward side of these mega-dunes is at most 300-400m, and usually much less, from the Western edge, i.e. base, of the dune (where you could walk with a magnetometer). The edge of the meteorite would be a further 100m into the dune today, making at the very most a total of up to 500m from the edge. From this distance the meteorite would be highly detectable by a magnetometer.

\begin{figure}[t]
    \centering
    \includegraphics[width=15cm]{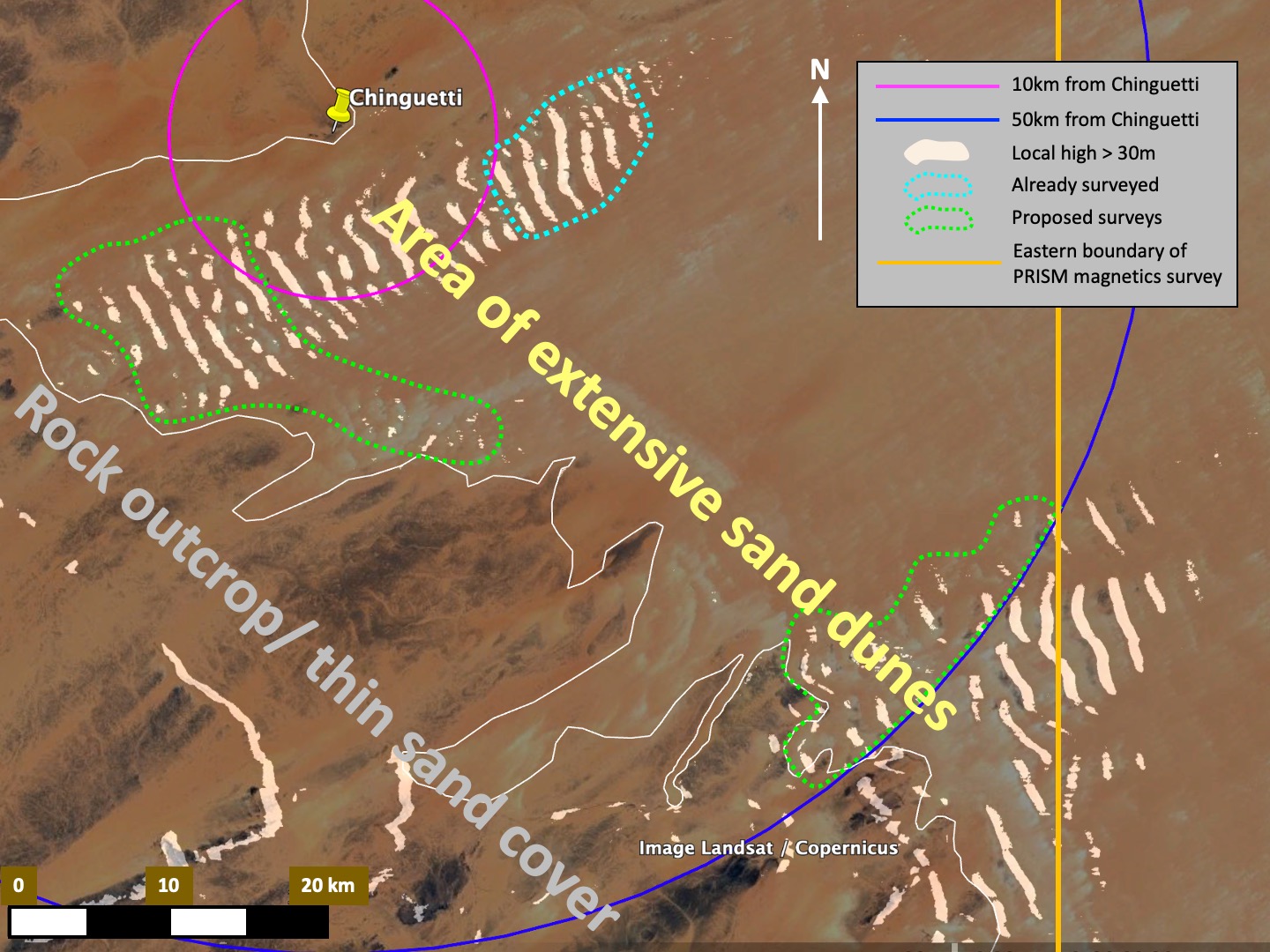}
    \caption{Map showing the high sand dunes, $>30$\,m height, to the south of Chinguetti.}
\end{figure}

We now justify the statement that the dunes can have drifted no more than 100\,m since Ripert's trip. The 8y baseline of the SRTM (early) and ALOS (late) DEM data allows us to estimate the rate of drift of the dunes.
Since the dune speed is very small, the DEM data maps at the two epochs are almost identical, but the SRTM data are noisier i.e. the uncertainties on the elevations are larger. Also the SRTM data include bad pixels which require masking. First we tried cross-correlation of the sections of the maps containing the dunes, accounting for the fact that the coordinate systems of the two datasets are offset by 0.5\,pix on each axis. We found that the cross-correlation results were unsatisfactory in that the measured shifts were sensitive to choices such as padding of the boundaries and whether to subtract the mean or median from the signal, or not to subtract at all. Instead, we measured shifts between the two datasets using straightforward least-squares fitting. A rectangle of a region of dunes in the SRTM data was selected, and bad pixels were masked. For a small grid of integer pixel shifts relative to the ALOS data the sum of the squares of the elevation differences between the datasets was computed at each shift point. Finally, an elliptical quadratic function, with 6 free parameters, including orientation, was fitted to this surface to measure the location of the minimum, at a shift of a fraction of a pixel. As a control we measured the apparent shifts in several rocky regions.

In the rocky regions the measured supposed drift speeds were consistent with zero, as required, with a scatter of 0.2\,m/yr. In the dunes there was some scatter between measurements of different areas, but the maximum speed measured was only 0.7\,m/yr. Therefore we conclude that a speed of 1\,m/yr represents an upper limit to the average drift of the dunes, based on this 8-year time span. Landsat (i.e. 2D) data exist over a much longer time span, with images separated by up to 36y. However, the older images are of lower resolution than more recent images, so marking, by eye, where an edge is located has uncertainty. Our estimates using these data-sets show speeds a little over 1.0\,m/yr.  

To summarize, points towards the W side of each stripe on Fig.1, marked `Local high $>30$\,m', and within the annulus 10\,km to 50\,km from Chinguetti, are the only feasible locations where the meteorite could be hidden. The western cliff edge of the meteorite should be located about 100m northeast of the western edge of one of these patches. The dotted areas, colored cyan, and green, are discussed below.

\section{The Fugro PRISM-I aeromagnetics data}
\label{sec:fugro}

Suitable aeromagnetic data exist over the area of interest. These data are in the custody of the Mauritanian Ministry of Petroleum Energy and Mines. The data acquisition and processing are described in \citet{Finn2015} and \citet{OConnor2005}, but the latter publication, which contains detailed maps, is also proprietary. \citet{Finn2015} state that the Fugro data were acquired as part of the PRISM-I project in one degree blocks. The aircraft was flown at a height of 100\,m with a line spacing of 500\,m. We understand from Finn (private communication) that the full coverage of the Fugro data is plotted in Fig. 8 of \citet{Finn2015}. Chinguetti is located within the ``Western'' area which includes longitudes 16W to 12W (in fact the Eastern limit is at about 11 56 W). So the one degree block of interest would cover 20N$<{\rm lat}<21$N 13W$>{\rm long}>12$W. In the PRISM-I data this block is labeled `Chinguetti 2012'. Happily, the eastern boundary, marked orange in Fig. 1, ensures that all the dunes at distances $<50$\,km are covered.

The data was acquired with credited funds primarily from the World Bank. Included in the objectives set by the World Bank was ensuring that the data acquired would be available for future use. So it is no surprise that the processed Fugro data have been made available to a number of teams for geological investigations recently, including \citet{Ba2020} (to study the Reguibat Shield), \citet{Aifa2021} (to model subsurface structures in the Tasiast area), and \citet{Abdeina2021} (to examine the Richat structure). \citet{Abdeina2021} provide a useful description of the processed data available which include 6 digital maps in each block: the total field magnetic anomaly, the reduced to the pole magnetic anomaly, the magnetic analytic signal, the magnetic horizontal gradient, and the first and second vertical derivatives.

In a letter to the Ministry dated 16 Jan 2023, and marked Confidential, we requested the PRISM aeromagnetic data in the Western area. We explained our scientific goals and the potential benefit to Mauritania. To our knowledge this is the first request of the data for the purposes of searching for the meteorite, and we would claim that this establishes priority. Despite frequent prompts we have yet to receive any data (Feb 2024). We wrote a second confidential letter, dated 16 December 2023, this time directly to the Minister, but no reply has been received. (We wrote to El Houssein Abdeina (at the University of Nouakchott, Mauritania) in November 2023 asking for assistance in obtaining the data, without stating our science goals. We specified the Western region, longitudes 18W to 12W. He kindly replied providing sight of the {\em total} magnetic field data over 18W to 13W, but not over the strip of interest 13W to 12W. We pointed out the omission, but have received no reply.) 

By examining the PRISM-I Fugro aeromagnetic maps (especially the reduced to the pole magnetic anomaly data) and searching for magnetic anomalies within the footprint of the dunes shown in Fig. 1 it should be possible to detect the giant Chinguetti meteorite, if it exists, or conversely rule out its existence. Magnetic anomalies outside the dune footprint should be treated as unrelated. All the dunes in Fig. 1 could be searched, but if several candidate anomalies are found, one should prioritize candidates in the 10 to 50\,km annulus, with lower priority accorded to candidates outside the annulus.

\section{Results from a small surface magnetometer survey} 

As discussed above, a surface magnetometer survey conducted by walking along the Western foot of each relevant dune, would be another way to answer with confidence the question of whether the meteorite exists. The eastern area of dunes marked cyan in Fig. 1 is of particular interest. As related in \citet{FerdeDieu}, in 1934 Captain Linar\`{e}s wrote to Monod recounting a conversation with a blacksmith who stated that the blacksmiths of Chinguetti recovered iron from a block `as big as my house' in a location to the E or SE of Chinguetti `roughly half way between Chinguetti and Ouadane'. Ouadane lies E of Chinguetti. The eastern limit of Les Boucles is certainly not half way to Ouadane, but there are no further large dunes to the E within our 50\,km limit.

Over the period 14 Dec to 17 Dec 2022 we conducted a surface-based magnetometer survey over the region marked in Fig. 1 enclosed by the dotted cyan line. We walked lines along the western edge of the six dunes, as well as the isolated patches visible. We walked continuous lines, taking readings approximately every 50m. The entire set of readings varied within a range of only 46 nT, which is of the same order of magnitude as diurnal variation, so the meteorite definitively does not lie in this region. 

Absent the PRISM-I data, to complete a useful surface magnetometer survey it is recommended to cover only dunes in the 10 to 50\,km annulus, which would mean surveying the green areas marked in Fig. 1. We estimate that to complete a survey of these areas would require a trip from Chinguetti lasting three weeks.

\section{Conclusion}

As described in \S~\ref{sec:fugro}, examination of the PRISM-I aeromagnetics data in the region south of Chinguetti, following the methodology described, can finally resolve the question of the existence of the Chinguetti meteorite in a definitive manner. If the result is negative the explanation of Ripert's story would remain unsolved, however, and the problems of the ductile needles, and the coincidental discovery of the mesosiderite would remain.

\section{Acknowledgements}

We are grateful to Amjad Moctar for assistance with many aspects of this project including the logistics of our magnetometer campaign, sourcing equipment, and acting as translator in Mauritania and as a general source of local knowledge. We thank Brigitte Zanda for filling in some historical details.

\bibliography{chinguetti}{}
\bibliographystyle{chicago}

\end{document}